\documentclass[a4paper,12pt]{spieman}  
\usepackage{amsmath,amsfonts,amssymb}
\usepackage{graphicx}
\usepackage{setspace}
\usepackage{tocloft}
\usepackage{soul}
\usepackage[nice]{nicefrac}
\usepackage{orcidlink}
\usepackage{hyperref}
\usepackage{lineno}
\usepackage{nicefrac}

\newcommand{\isro}{{\it ISRO}}
\newcommand\suit{{\it SUIT~}}
\newcommand{\degree}{$^{\circ}$}
\newcommand{\rv}[1]{{\color{black} #1}}
\newcommand{\js}[1]{{\color{black} #1}}

\graphicspath{{./}{pics/}}

\cftpagenumbersoff{figure}
\cftpagenumbersoff{table}

\title{Photometric Calibration \& Spectral Validation of the Solar Ultraviolet Imaging Telescope onboard Aditya-L1}

\author[a,b,*]{Janmejoy Sarkar \orcidlink{0000-0002-8560-318X}}
\author[a]{Soumya Roy \orcidlink{0000-0003-2215-7810}}
\author[a]{A N Ramaprakash \orcidlink{0000-0001-5707-4965}}
\author[a]{Rushikesh Deogaonkar \orcidlink{0009-0000-2781-9276}}
\author[c,*]{Sreejith Padinhatteeri \orcidlink{0000-0002-7276-4670}}
\author[a]{Durgesh Tripathi \orcidlink{0000-0003-1689-6254}}
\author[a,d]{Avyarthana Ghosh \orcidlink{0000-0002-7184-8004}}
\author[e]{Raja Bayanna Ankala \orcidlink{0000-0001-5802-7677}}
\author[b]{Gazi Ameen Ahmed \orcidlink{0000-0002-0631-4831}}

\affil[a]{Inter-University Centre for Astronomy and Astrophysics, Post Bag 4, Ganeshkhind, Pune 411007}
\affil[b]{Tezpur University, Napaam, Tezpur, Assam 784028}
\affil[c]{Manipal Centre for Natural Sciences, Manipal Academy of Higher Education, Manipal - 576104, India}
\affil[d]{Now at Embedded Devices \& Intelligent Systems, TCS Research, India}
\affil[e]{Udaipur Solar Observatory, Udaipur.}
\begin{document} 
\maketitle
\begin{abstract}
The Solar Ultraviolet Imaging Telescope (\suit) is one of the seven payloads on board Aditya-L1 mission of the Indian Space Research Organization (\isro). \suit provides full and partial disk images of the Sun in the 200{--}400 nm wavelength range. This would help us probe the solar atmosphere at different heights and understand the mass and energy transfer process between its layers. 
For the first time, \suit will also help us measure spatially resolved solar spectral irradiance at this wavelength band, which is significant for studying the sun-climate relationships.
To perform these studies, it is necessary to photometrically calibrate the payload and validate the spectral coverage of the various bandpasses.
We perform the photometric calibration and spectral validation of 8 bandpasses using light of known intensity and spectral coverage. 
For photometric calibration, the telescope throughput is modeled using sun-as-a-star spectrum from SOLSTICE and SOLSPEC. The modeled throughput is compared with in-lab measurements taken with light of known intensity. The ratio of measured photoelectrons gathered with the modeled prediction agree within \rv{20\%}.
For spectral validation, readings are taken across the transmission spectrum of each
filter, keeping adjacent readings independent of each other. The relative intensity measured at each wavelength is seen to trace the modeled telescope bandpass for that filter.
These tests could not be performed for filters with bandpasses operating below 250 nm (NB01, BB01 and BB02), primarily due to heavy atmospheric attenuation in these wavelengths leading to decreased SNR of the data.
The experimentally measured results agree closely with the modeled values, validating \suit's optical performance and presenting the reliability of the developed throughput model.
\end{abstract}


{\noindent \footnotesize\textbf{*} Janmejoy Sarkar,  \linkable{janmejoy.sarkar@iucaa.in} }
{\noindent \footnotesize\textbf{*} Sreejith Padinhatteeri,  \linkable{sreejith.p@manipal.edu} }


\section{Introduction}\label{sec:intro}

The Solar Ultraviolet Imaging Telescope \suit \cite{ghosh16,suit17} is an instrument on-board Aditya{--}L1 mission \cite{adityal1,aditya} of the Indian Space Research Organization (\isro) launched on September 2, 2023. \suit provides images of the Sun in the near ultraviolet (NUV) range (200{--}400 nm), using eleven filter combinations. SUIT is designed to record full-disk and partial-disk images of the Sun with a spatial resolution of 1.4". 
\js{These wavelengths mainly originate from the solar photosphere and chromosphere, thereby helping us understand the dynamic coupling between these magnetized layers. This wavelength band is also crucial to measure and monitor the spatially resolved solar spectral irradiance which is central to the chemistry of oxygen and ozone in the Earth's stratosphere \cite{haigh07}. Additionally, this high-resolution imaging capability enables a detailed study of NUV irradiance from solar flares, which might hold key indications of early flare signatures \cite{panos20}. Moreover, combining flare observations from these images with those recorded in EUV and X-rays will help us develop a complete spectral energy distribution in solar flare, which is crucial to understand the energetics in these dynamic events. 
To date, there have been no full-disk observations in this band of 200-400 nm (except in Ca II k from ground-based instruments). The Ultraviolet Spectrometer and Polarimeter (UVSP) \cite{uvsp} on the Solar Maximum Mission (SMM) \cite{smm}, and the Interface Region Imaging Spectrograph (IRIS) \cite{iris} have had limited spatial and spectral coverage in these wavelengths. IRIS records full disk mosaics of the Sun, which cannot give simultaneous imaging of the entire solar disk, like \suit.
The absolute throughput of IRIS is derived from the efficiency curves of each individual optical element. Post-launch, IRIS observes B-type stars or compares the full-disk mosaic measurements with co-temporal sun-as-a-star SORCE/SOLSTICE or TIMED/SEE observations \cite{iris_calib}. 

On the other hand, the Sunrise Filter Imager (SuFI) \cite{sufi_calib} instrument does cover this wavelength band in two flights of the SUNRISE balloon-borne observatory, but with a very small field of view and extremely high resolution \cite{sunrise_barthol, sunrise_2, sunrise_sami} ($\approx$~0.02 arcsec/pixel). On-ground contrast transfer, spatial resolution tests, and distortion tests were performed for this instrument \cite{sufi_calib}. However, SuFI did not undergo any on-ground or on-board photometric calibration. The instrument measured relative contrast values at extremely small spatial scales, but not in the absolute photon fluxes. The Sunrise Ultraviolet Spectrograph and Imager \cite{susi} instrument also covered 309~nm in the third flight of Sunrise, but at extremely high resolution and a small field of view.
On-ground photometric and spectral calibration of full-disk solar telescopes working in the same wavelength band as \suit is limited. This paper provides a complete account of the calibration procedure while discussing the experimental setup and contamination control in detail for \suit.
}

\begin{figure}[ht!]
\begin{center}
\begin{tabular}{c}
\includegraphics[width=1\linewidth]{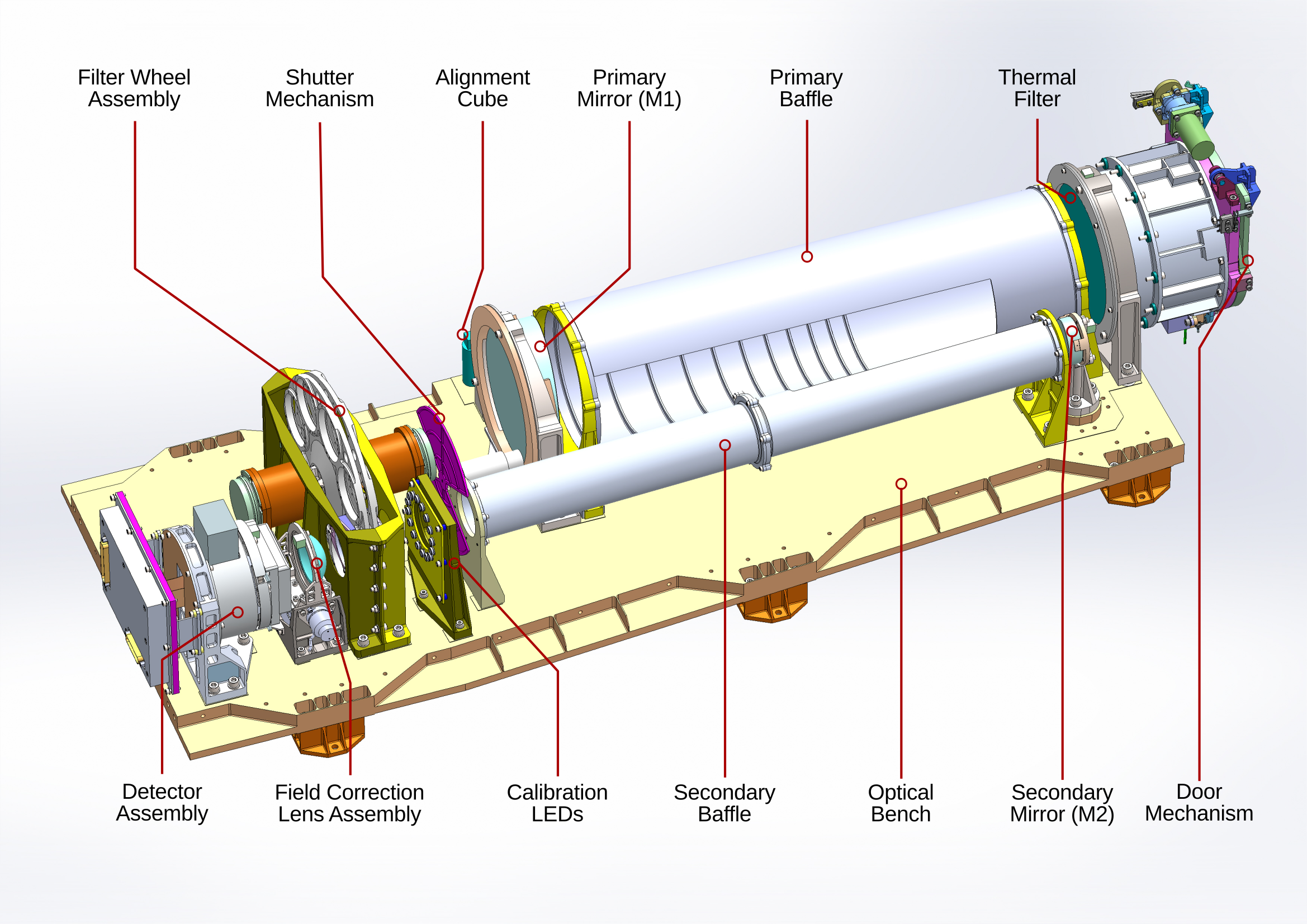}
\end{tabular}
\end{center}
\caption 
{ \label{fig:suit} A schematic diagram of the \suit payload. } 
\end{figure}

\suit consists of two main sub-units: The \suit optics package- comprising of an off-axis Ritchey-Chr\'{e}tien telescope, and \suit Electronics package- responsible for communicating with the telescope to perform imaging. The main components of the SUIT optics package consist of a multi-operation entrance door, a thermal filter to control the amount of incoming sunlight, primary and secondary mirrors, a shutter mechanism to control exposure times, baffles to reduce stray and scattered light, a motorized filter wheel assembly, a piezoelectric focusing mechanism, and a CCD detector. Fig \ref{fig:suit} shows a schematic diagram of the telescope.

\begin{figure}[ht]
\begin{center}
\begin{tabular}{c}
\includegraphics[trim={2cm 0.8cm 3cm 2.4cm},clip,width=0.7\linewidth]{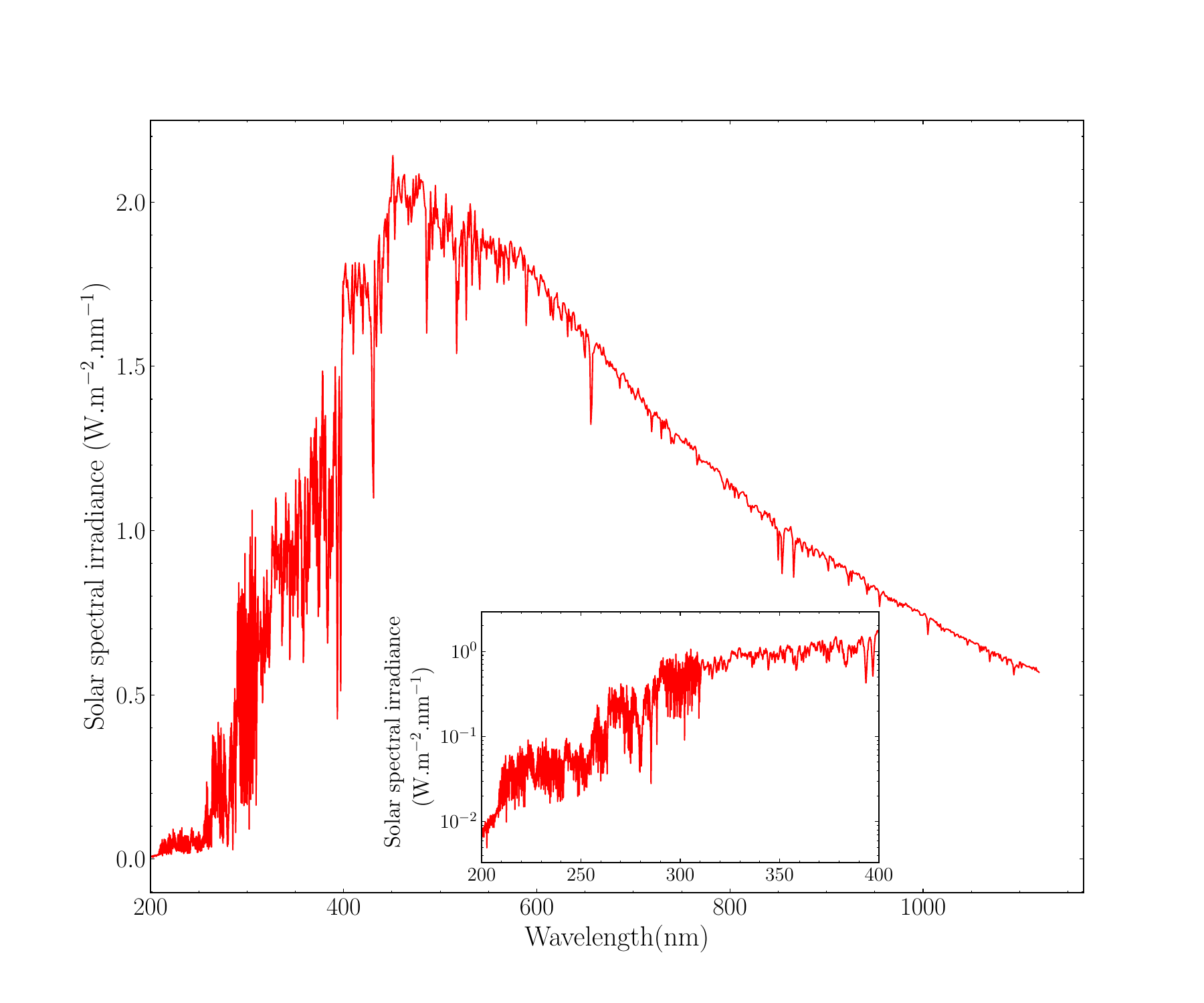}
\end{tabular}
\end{center}
\caption {The solar spectrum from SOLSTICE and SOLSPEC combined. The inset plot shows a \rv{$\sim$ two orders of magnitude} rise in solar irradiance within \suit's observation band.} 
\label{fig:sun_spec} 
\end{figure}

The solar flux increases by two orders of magnitude, in the wavelength range of interest (200 {--} 400 nm). Figure ~\ref{fig:sun_spec} shows \rv{the concatenated spectra from SOLar STellar Irradiance Comparison Experiment (SOLSTICE; 115{--}320 nm) onboard the SOlar Radiation and Climate Experiment (SORCE) \cite{rottman05,harder05,mcclintock05} satellite and SOLar SPECtrometer (SOLSPEC; $> 320$ nm) \cite{thuillier09}.}
\rv{To accommodate this large dynamic range and get sufficient SNR in the observations, we use specific combination filters with each bandpass filter. \suit has 8 narrowband (NB) filters, 3 broadband (BB) filters, and 5 combination filters mounted on two filter wheels, each having eight slots. A given filter combination is achieved by rotating the two filter wheels independently and placing the desired filter combination in the beam path\cite{suit17}.} The filter combinations are listed in Table~\ref{tab:science_filters}.
\js{The beam passing through the filters has an f-ratio of $\sim f/25$. For any point on the Sun, the diameter of the spot size at the filter is $\sim 4$ mm. The spatial characterization of each filter shows negligible change in the central wavelength and bandwidth. Moreover, filters showing the least spatial variation is chosen from a lot, thereby ensuring uniform illumination across the area of the filter.}
Note that \suit uses BB01 as a combination filter for NB01 and BB01. Similarly, NB08 is used as a combination filter for another NB08 filter. More details on \suit filters and their characterization are elaborated in Sarkar et. al.\cite{scfilt}.
 
\begin{table*}[ht]
\caption{List of science filters on board \suit. Columns (from left to right) denote filter mnemonics (including science and combination filters; NB: Narrowband, BB: Broadband, BP: Bandpass), central wavelengths for science filters, corresponding bandpasses, and the observation interest for the filter.} 
\label{tab:science_filters}
\begin{center}
\begin{tabular}{||l|c|c|c|r||}
\hline
\textbf{Science}  &	\textbf{Combination} &	\textbf{Central} & \textbf{Bandpass} &\textbf{Science} \\
\textbf{Filter}	&	\textbf{Filter}     &	\textbf{Wavelength  (nm)}	&		\textbf{(nm)	}	   	&\textbf{target}		\\
\hline
NB01    & BB01 		& \rv{214.0} 		& 11.0 		& Continuum\\
NB02 	& BP02		& \rv{276.7}			& 0.40 		& Mg~\rm{II}~k blue wing \\
NB03 	& BP02		& \rv{279.6} 		& 0.40 		& Mg~\rm{II}~k\\
NB04 	& BP02		& \rv{280.3}			& 0.40 		& Mg~\rm{II}~h\\
NB05	& BP02		& \rv{283.2}		    & 0.40 		& Mg~\rm{II}~h red wing\\
NB06 	& BP03		& \rv{300.0} 		& 1.00 		& Continuum\\
NB07 	& BP03		& \rv{388.0}			& 1.00 		& CN Band\\
NB08	& NB08		& \rv{396.9} 		& 0.10 		& Ca~\rm{II}~h\\
BB01 	& BB01		& \rv{220.0}			& 40.0		& Herzberg Continuum \\
BB02 	& BP04		& \rv{277.0} 		& 58.0      & Hartley Band\\
BB03 	& BP04		& \rv{340.0}			& 40.0      & Huggins Band\\
\hline
\end{tabular}
\end{center}
\end{table*}

\rv{SUIT is designed to deliver relative photometric measurements within 1\% accuracy from one spatial location to another in an image. To achieve the scientific goals, photometric and spectral calibration of the payload is of utmost importance.}
The model details and derived design values are discussed in \S\ref{sec:model}. The telescope is calibrated for throughput accuracy during the final test before integration on the satellite. Details of the calibration setup are explained in \S\ref{sec:setup}. We present the results from in-lab calibration and its comparison with the design values in \S\ref{sec:calibration} and \S\ref{valid}. Finally, we summarize and conclude in \S\ref{sec:conc}.

\section{Telescope Throughput Model}\label{sec:model}

\suit telescope design is performed considering the requirement of throughput and photometric accuracy. Optical parameters of different optical components, like reflectivity of the primary and secondary mirrors, thermal filter transmission, field corrector lens transmission, spectral transmission of the science filter, and the combination filter, are tuned to deliver the best possible throughput at corresponding wavelength ranges. 

\begin{figure}[h!]
    \centering
    \includegraphics[width=0.9\textwidth]{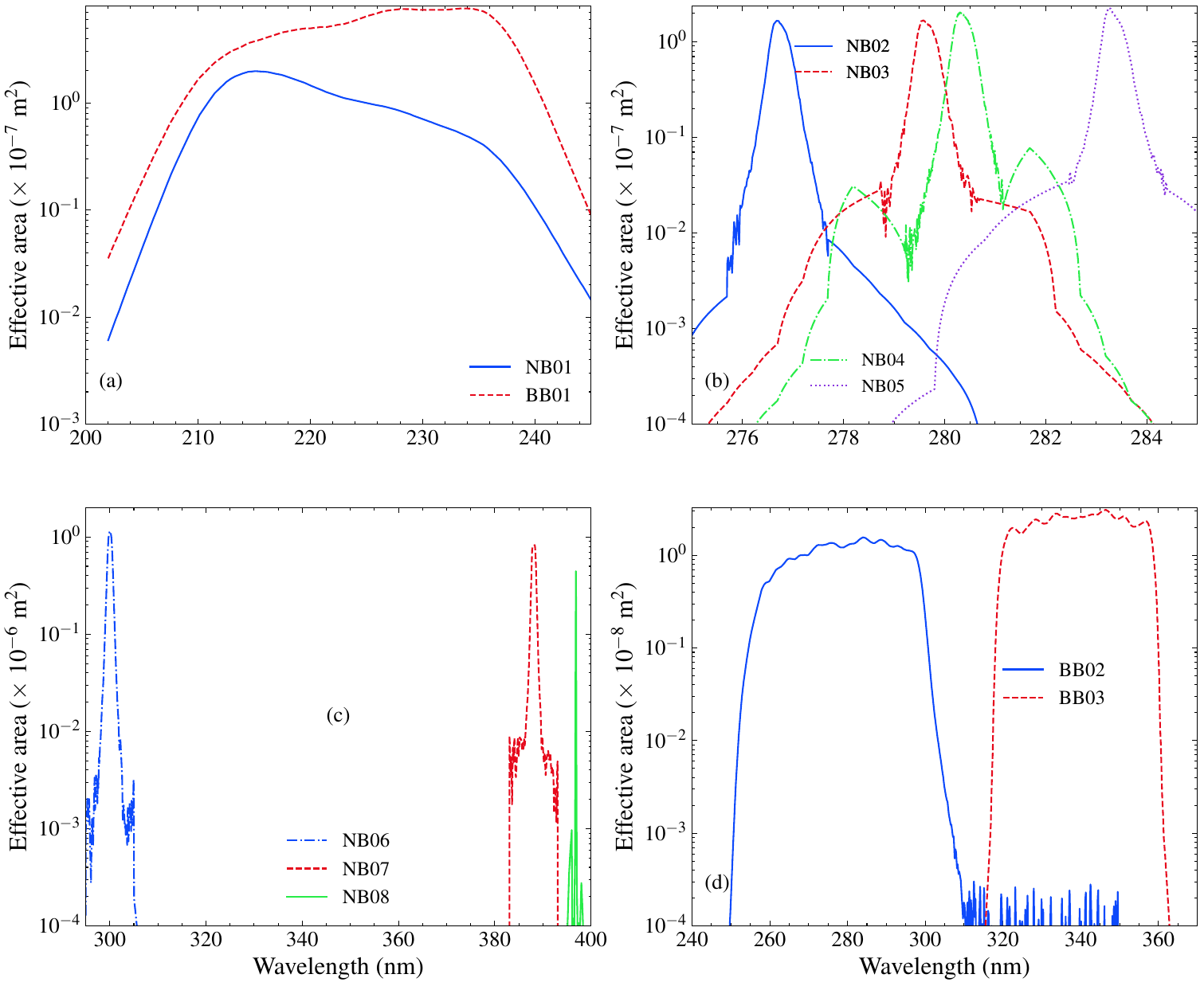}
    \caption{The effective area curves for SUIT science filters. \rv{The central wavelengths and the observation interest for each bandpass are tabulated in Table \ref{tab:science_filters}}.}
    \label{fig:eff_area}
\end{figure}

\rv{The end-to-end optical response of the instrument is modeled from the lab-measured parameters (transmission, reflectivity, quantum efficiency) of each component using Equation \ref{eq1}. We use the concatenated solar spectral irradiance from SORCE-SOLSTICE and SOLSPEC, along with the optical response, to predict the expected signal. Note that in this wavelength range ($\mathrm{> 320~nm}$), SOLSPEC offers the best available wavelength resolution measurements, and is therefore used for our work. The concatenated spectrum is interpolated to the same spectral resolution as that of the corresponding science filter spectral measurements. This is necessary to decide the exposure times of the payload in the eleven science bandpasses for various modes of operation.
 
 \begin{equation}\label{eq1}
     \mathrm{DN~=~\int_{\lambda_1}^{\lambda_2} ~P(\lambda)~R(\lambda)~t~d\lambda}
 \end{equation}

Here,
\begin{itemize}
    \item $\mathrm{P(\lambda)}$= Incoming photon flux ($\mathrm{Photons~m^{-2} s^{-1}}$) from the Sun falling on the entrance aperture of SUIT. The total number of photons is integrated over a wavelength range of 200 nm to 1000 nm. This is derived from the solar spectral irradiance (SSI) from the concatenated SORCE-SOLSTICE and SOLSPEC spectra in units of $Wm^{-2}nm^{-1}$. The SSI is divided by photon energy ($hc/\lambda$) of the corresponding wavelength to derive the number of incident photons for that particular wavelength bin. This is in units of the number of photons.

    \item $\mathrm{t}$= Exposure Time in seconds.

    \item $\mathrm{R(\lambda)}$= Effective Area of the end-to-end configuration of SUIT for a particular science filter combination. This is derived by multiplying the measured transmittance/reflectance of all the optical sub-assemblies in the beam path, along with the sensitivity and analog to digital conversion of the CCD and its electronics. This can be expressed as Equation \ref{eq:eff_area}.
    \begin{equation}
        \mathrm{R(\lambda)~=~TF(\lambda)\times PM(\lambda)\times SM(\lambda)\times SF_{i}(\lambda)\times SF_{j}(\lambda)\times L(\lambda)\times QE(\lambda)\times G \times A}
        \label{eq:eff_area}
    \end{equation}    
    where the measured dimensionless spectral profiles, and the corresponding spectral step size (in parentheses) are given as,
    \begin{itemize}
        \item $\mathrm{TF(\lambda)}$= Thermal filter at entrance aperture (5 nm).
        \item $\mathrm{SF_{i}(\lambda) ~ and ~ SF_{j}(\lambda)}$= Science filters ({0.01{--} 0.02 nm for various filters}).
        \item $\mathrm{L(\lambda)}$ = Field correction lens (5 nm).
        \item $\mathrm{PM(\lambda)~and~SM(\lambda)}$= Primary and Secondary mirrors (1 nm).
        \item $\mathrm{QE(\lambda):}$ CCD quantum efficiency (Photoelectrons: $pe^-$).
        \item $G$= CCD electronics gain ($\nicefrac{DN}{pe^-}$)
        \item $\mathrm{A}$= Area of the telescope aperture in $\mathrm{m^2}$.
    \end{itemize}
    For further operations, the quantities are interpolated to the smallest spectral sampling rate of all measurements so as to resolve various spectral features observed to be observed by \suit.
\end{itemize}
The incoming rate of photon flux ($\mathrm{P(\lambda)}$) is multiplied by the telescope response function ($\mathrm{R(\lambda)}$) and exposure time. This is integrated over the wavelength band of interest (between $\mathrm{\lambda_1}$ and $\mathrm{\lambda_2}$) to simulate the data numbers detected by SUIT in a particular filter combination. The effective area curves for each filter are shown in Fig.~\ref{fig:eff_area}.}

\section{Experimental Setup}\label{sec:setup}
\rv{An experimental setup is developed, as illustrated in Figure \ref{fig:experimentalsetup}, to measure the end-to-end throughput of the instrument and compare the measured results with the throughput model described in \S \ref{sec:model}.} Light of known \rv{spectral irradiance} is fed into the payload from a monochromator after collimation. The payload is operated in a vacuum chamber to simulate space-like conditions in this setup. 

\subsection{Contamination control and payload preparation}
Given the operational wavelength range, the throughput of \suit is highly susceptible to particulate and molecular contamination. Therefore, it is assembled and stored in a Class 100 clean room with constant purging of 5.0 purity gaseous nitrogen. 
In space, the SUIT CCD is passively cooled to {--}55{\degree}~C with a radiator plate and a cold finger. To attain the same level of cooling in lab, the payload is operated in a contamination-controlled 0.7~m custom-made vacuum chamber at a pressure of $< 10^{-6}$~mbar. Liquid nitrogen is fed by closed-loop micro-dosing through a cooling jacket mounted over the CCD heat pipe, instead of using the radiator plate.

Before loading the payload, the vacuum chamber is cleaned with isopropyl alcohol and baked at  $70^\circ $C for 24 hours. Temperature Controlled Quartz Crystal Microbalance (TQCM) verification is performed with the crystal at -10{\degree}~C in high vacuum.
Frequency change of $\leq 3~\mathrm{Hz/hr}$ over 24 hours was considered satisfactory for operating \suit in the chamber. However, in practice $< 2~\mathrm{Hz/hr}$ was achieved over 36 hours. This corresponds to a contaminant deposition rate of $3.92 \times 10^{-9}~\mathrm{g/cm^2/Hr}$ on the TQCM crystal ($1 Hz / hr \equiv 1.96 \times 10^{-9} \mathrm{g/cm^2/hr}$ ).

The photometric and spectral calibration of \suit is performed after payload vibration, acoustic, and thermovacuum qualification tests. This required transportation of the payload to various clean rooms and labs. To avert any risk of contaminating the CCD upon cooling, the payload is subjected to a bake-out at 40{\degree}~C for 36 hours after loading in the chamber. TQCM verification is ensured to be $<~3~\mathrm{Hz/hr}$ over 24 hours after payload baking.

\subsection{SUIT Collimator}\label{sec:collimator}

\begin{figure*}[ht]
\includegraphics[width=1.0\linewidth]{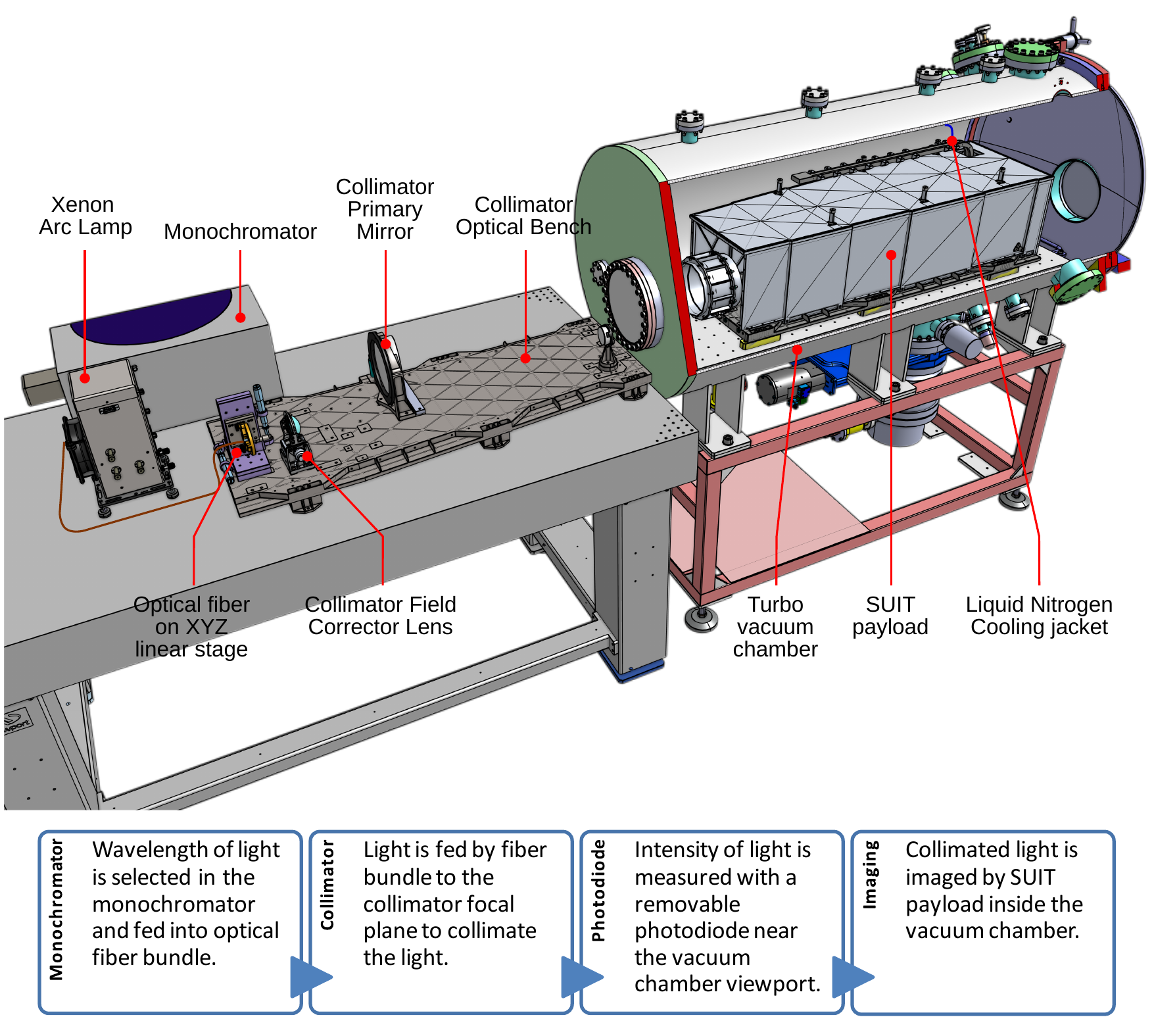}
\caption 
{\rv{Diagram [top] and schematic flowchart [bottom] for photometric calibration and spectral validation of \suit showing the section view of the turbo vacuum chamber.}} 
\label{fig:experimentalsetup}
\end{figure*}

The \suit flight spare is used as a collimator because of comparable optical performance as the payload in NUV. An illuminated target is kept at the focal plane, light from which emerges collimated from the front aperture.
The \suit collimator exhibits an RMS wavefront error of 41.4~nm ($\sim \lambda/7 ~ for ~ 300~ nm$) as compared to the \suit payload's 34~nm ($\sim \lambda/9 ~ for ~ 300~ nm$). The Zernike coefficients of the collimator and the payload are tabulated in Table \ref{table: zernikes}.
The collimator is mounted on an optical table so that the front aperture is aligned with the viewport of the vacuum chamber housing the SUIT payload as in Fig \ref{fig:experimentalsetup}.
The optical bench is designed to rest on six pods bolted to the optical table. The telescope and collimator are of RC design, causing the wavefront error to change grossly in case of stress-induced misalignment. Therefore, all bolts are uniformly and gradually torqued to 400 N-cm to avoid mechanical distortion of the optical bench while ensuring planarity with a feeler gauge and filling any gaps between the pods and the bench with appropriate shims. 

A big flat mirror (dia: 150 mm) is placed at the front aperture of the collimator along the payload optical axis. The collimator's wavefront error is verified with a Fizeau Interferometer by focusing an f/25 beam at the focal plane and retro-reflecting it from the reference mirror at the front aperture, to ensure proper mounting.
Following this, an f/3.5 transmission sphere is used with the interferometer to overfill the mirrors of \suit collimator. Repeated interferograms are taken, and the fringes are constantly monitored while the interferometer is moved along the optic axis. This is done till the fringes appear straight instead of circular, indicating nullification of power. The point of convergence of the beam from the interferometer is the focal point of the collimator, indicating the location for a target.
\begin{table*}[ht]
\caption{Zernike coefficients representing the extent of optical aberrations of the \suit payload and collimator.}
\label{table: zernikes}
    \begin{center}
			\begin{tabular}{|c|c|c|}
				\hline
				\rule[-1ex]{0pt}{2.5ex} \textbf{Aberration} & \textbf{\textit{SUIT} payload (nm)} & \textbf{Collimator (nm)}\\
				\hline
				\hline
				\rule[-1ex]{0pt}{2.5ex} \textbf{RMS} & 34 & 41.4 \\
				\hline
				\rule[-1ex]{0pt}{2.5ex} \textbf{PVr} & 196.4 & 210.3 \\
				\hline
				\rule[-1ex]{0pt}{2.5ex} \textbf{Power} & 25 & -4.4 \\
				\hline
				\rule[-1ex]{0pt}{2.5ex} \textbf{Astig X} & -32.7 & 36.7 \\
				\hline
				\rule[-1ex]{0pt}{2.5ex} \textbf{Astig Y} & 22.8 & 0.88 \\
				\hline
				\rule[-1ex]{0pt}{2.5ex} \textbf{Coma X} & -30.8 & -63.3 \\
				\hline
				\rule[-1ex]{0pt}{2.5ex} \textbf{Coma Y} & -9.9 & 28.8 \\
				\hline
				\rule[-1ex]{0pt}{2.5ex} \textbf{Primary Spherical} & -19.8 & -18.8 \\
				\hline
				\rule[-1ex]{0pt}{2.5ex} \textbf{Trefoil X} & -13.9 & -40.4 \\
				\hline
				\rule[-1ex]{0pt}{2.5ex} \textbf{Trefoil Y }& -17.3 & -19.3 \\
				\hline
			\end{tabular}
		\end{center}
	\end{table*} 

Light from the collimator is fed into the vacuum chamber \rv{through an Allectra Deep UV Viewport \js{(110-VPQZ-C200-DUV) which provides $\mathrm{> 99.8\%}$ internal transmission at 248 nm, ensuring nearly complete transmission inside the glass. However, $\sim 4\%$ of the light reaching each surface is lost due to reflection.}}
The light entering the chamber is focussed and imaged by \suit. To avoid any light loss, aligning the lateral position and tilt of the payload with the collimator is necessary. This intricate alignment of the collimator and the payload is achieved with a theodolite, which also functions as an autocollimator.

The theodolite is autocollimated with the Collimator axis using the reference mirror. With assistance from repeater mirrors and theodolite, the axes of the collimator are matched with that of the payload. The optical axis of \suit is given by a reference cube placed on the optical bench, outside the cover panel. The payload is finely shifted within the vacuum chamber to achieve autocollimation with the alignment cube. This aligns the payload in the vacuum chamber with the collimator facing it outside the vacuum chamber.

\subsection{Monochromator configuration and calibration} \label{sec:monochromator_config}

For photometric and spectral calibration, collimated light from a known wavelength band and intensity is fed into the payload. For this purpose, we use Andor Shamrock 500i spectroscope/ monochromator with an optical fibre (Andor SR-OPT-8024). One end of the optical fiber bundle forms a linear array with 19 fibers of 0.2 mm thickness. This end is attached to the monochromator and behaves like an exit slit with a thickness of 0.2~mm and height 4.655~mm. The other end of the fibre forms a circular bundle and is placed at the focal plane of the collimator. 

A Mercury light source (Ocean Optics HG2 Wavelength Calibration Source) is used for wavelength calibration of the monochromator. The Mercury lamp is kept at the entrance aperture, and the monochromator is set to observe a relatively isolated spectral line at 253.65 nm. Using a micrometer actuator, the position of the fiber slit is adjusted to get maximum intensity at the fiber face. This is validated with another relatively isolated line (365 nm) within the 200{--}400~nm band.
The mercury lamp is then replaced with a Xenon Arc Lamp (75W Newport 60000 Q Series Lamp with stabilized power supply), which is focused at the input slit of the monochromator. 

\section{Photometric Calibration}\label{sec:calibration}

In this test, an optical fiber is used to feed light from the monochromator at the focal plane of the SUIT collimator, as shown in Fig \ref{fig:experimentalsetup}. The monochromator uses 1200 lines per mm (lpmm) grating with a nominal dispersion of 1.44 nm/mm, blazed at 500 nm. The spectral resolution of the monochromator depends on the entrance slit or the exit slit, whichever is geometrically wider. In our case, an entrance slit width of 3~mm is used to maximize the light output from the \rv{spectroscope, delivering wavelength band of 4.32~nm. This wavelength band accommodates the complete transmission spectrum of all the narrowband science filters (Table \ref{tab:science_filters}) being tested.}
	
A UV-enhanced National Institute of Science and Technology (NIST) traceable photodiode (Newport 918D-UV-OD3R) is kept at the exit aperture of the collimator, \rv{ which measures the net collimated output flux (in $nW cm^{-2}$) falling on the photodiode with area $1 cm^2$. An average of 150 independent readings are taken in a time span of 10 seconds.} The monochromator's and photodiode's wavelengths are set based on the science filter being tested. 

\rv{The power measured with the photodiode (nW) is normalized by the collecting area ($\mathrm{nWm^{-2}}$) and divided by the wavelength band of the spectrograph to get the spectral flux from the collimator (in $\mathrm{W m^{-2} nm^{-1}}$). The xenon light intensity changes linearly within this wavelength band, making it viable to divide the photodiode output by 4.32 nm to get the spectral flux.}
Also, the incoming wavelength band is wide enough to accommodate the entire transmission spectrum of the narrowband science filters (typically $< 1~nm$ FWHM). Therefore, the wavelength band incident from the monochromator is sufficient to perform the photometric test with the SUIT payload. The photodiode reading is recorded. 
Following this, the fiber face at the collimator is covered, and the background value seen on the photodiode is recorded. Finally, the photodiode is removed from the beam path, and an image of the fiber bundle is taken with \suit. The fiber face is covered to record the background value, and another exposure is taken. These results can be directly compared with the modeled throughput derived from the sun-as-a-star spectrum for various filter wavelength bands.
\js{The measurements are not affected by out-of-band transmissions, such as those arising from higher grating orders. The \suit science filters transmit only 0.01\% - 0.001\%  out-of-band radiation, significantly cutting out wavelengths beyond the desired bandpass (Table \ref{tab:science_filters}). In addition, our longest test wavelength is $\sim 396.9 nm$. The second order contamination, if at all, would arise from $\sim 200 nm$. Since we operate the collimator in air, these higher-order wavelengths are heavily attenuated. We were unable to detect emission below $\sim 250 nm$, even in the first order.}

The setup illustrated in Fig.~\ref{fig:experimentalsetup} is employed for the test. \rv{We can predict the solar throughput in various passbands from the experiment results and compare them with the modeled throughput derived from the sun-as-a-star solar spectrum using,

  \begin{align}
    & \mathrm{R= \dfrac{\dfrac{DN_{0.99}}{E_{0.99}}\times E_\odot}{DN_{simulated}}~where},
    \label{eqn:count_pred}
  \end{align}

  \begin{align*}
    \mathrm{DN_{0.99}} =&~\mathrm{Data~number~per~second~within~99\%~enclosed~counts}\\ 
                        & \mathrm{raidus~of~the~fiber~image~(In~units~of~DN.s^{-1})} \\
    \mathrm{E_{0.99}} = &~\dfrac{Photodiode~Output~[nWm^{-2}]\js{*0.92}*\Delta\lambda [nm]}{4.32~nm} ~\mathrm{(In~units~of~nW.m^{-2})}\\  
                        &~\mathrm{Measured~input~flux~from~the~photodiode~in~the~bandpass~[\Delta\lambda]}\\
                        &~\mathrm{of~the~concerned~filter.\js{~Corrected~for~reflective~losses~at~viewport~surfaces.}}\\
    \mathrm{E_{\odot}} =& ~\int_{\lambda_{1}}^{\lambda_{2}}~P(\lambda)~d\lambda ~(In~units~of~nW.m^{-2}) \\
                        &~\mathrm{Incident~energy~from~a~sun~as~a~star~spectrum}\\ 
                        &~\mathrm{in~the~concerned~filter's~passband.}\\ 
    \mathrm{DN_{simulated}} =& ~\int_{\lambda_{1}}^{\lambda_{2}}~P(\lambda)~R(\lambda)~t~d\lambda ~(In~units~of~DN.s^{-1})\\
                             &~\mathrm{Predicted~data~number~from~a~sun~as~a~star}\\
                             &~\mathrm{spectrum~using~the~throughput~model}\\           
  \end{align*}

\noindent We calculated the quantities in Eqn.~\ref{eqn:count_pred} using the following steps:

\begin{itemize}
    \item The background count is estimated by taking the median of the fiber occulted image.
    \item From the background corrected optical fiber images, we calculate the total counts within 99\% enclosed counts radius. We use 99\% to ensure we collect most of the light from the fiber ($\mathrm{DN_{0.99}}$).
    \item We use the concatenated SOLar STellar Irradiance Comparison Experiment (SOLSTICE; 115{--}320 nm) onboard the SOlar Radiation and Climate Experiment (SORCE) \cite{rottman05,harder05,mcclintock05} satellite and SOLar SPECtrometer (SOLSPEC; $\mathrm{>320 nm}$) \cite{thuillier09} solar spectrum as our standard for solar radiation. Using the spectra, we calculate the total energy input from the Sun for the concerned filter within the wavelength extent, $\mathrm{E_{\odot}}$.
    
    \item The photodiode gives the measurement of the corresponding input flux for an input spectral bin of 4.32 nm. This is used to calculate the input flux in the bandpass of the concerned filter,
    \js{
    $$\mathrm{E_{0.99}(\mathrm{in~units~of~nW.m^{-2}})}= \dfrac{Photodiode~Output~[nWm^{-2}]\js{*0.92}*\Delta\lambda [nm]}{4.32~nm}$$
    $\sim 4\%$ of the flux reaching each viewport surface is lost to reflection. This is accounted for by the factor of $\sim 0.92$.}
    
    \item Using Eqn.\ref{eq1} along with the aforementioned spectrum, we can now calculate the expected counts observed (i.e., $\mathrm{DN_{simulated}(in~units~of~DN.s^{-1})}$ for a given filter combination.
\end{itemize}
}

In Table~\ref{tab:throughput}, we list the comparison between the solar counts inferred from our measurements and the counts calculated from the throughput model for some of the filter combinations. Note that NB01, BB01, and BB02 have transmission below 250~nm. The atmosphere heavily attenuates UV light below this wavelength while traversing through the monochromator and collimator, making the signal insufficient for measurement. Also, the Xenon light source radiance increases almost linearly by 60 times between 200~nm and 300~nm, beyond which the \rv{emission} is relatively flat till 400~nm. This leads to insufficient SNR in the output during photometric calibration and spectral validation for these filters.

Our results demonstrate that the measured data counts per second are within 20\% of those derived from the throughput model (refer Table \ref{tab:throughput}), except for NB07. A reason for this can be traced to using SOLSPEC data to model the throughput for wavelengths beyond 320~nm and SOLSTICE data for modeling throughputs below 320~nm. SOLSPEC has a resolution of 0.5~nm, unlike SOLSTICE data offering a resolution of 0.1~nm.
This leads to lesser accuracy of the modeled data counts for filters operating beyond 320 nm. The modeled counts are not drastically different for NB08, despite lying beyond 320 nm. This can be attributed to the extremely narrow (\rv{0.1 nm}) bandpass of NB08 centered at the prominent Ca II h spectral line. BB03, on the other hand, has an effective bandpass of 40 nm, which makes the model oblivious to small-scale fluctuations in the sparsely sampled SOLSPEC solar spectrum and yields similar data numbers per second to that measured in the lab. 

\begin{figure}[ht!]
    \centering
    \includegraphics[trim={3cm 0.5cm 2cm 2cm},clip,width=\textwidth]{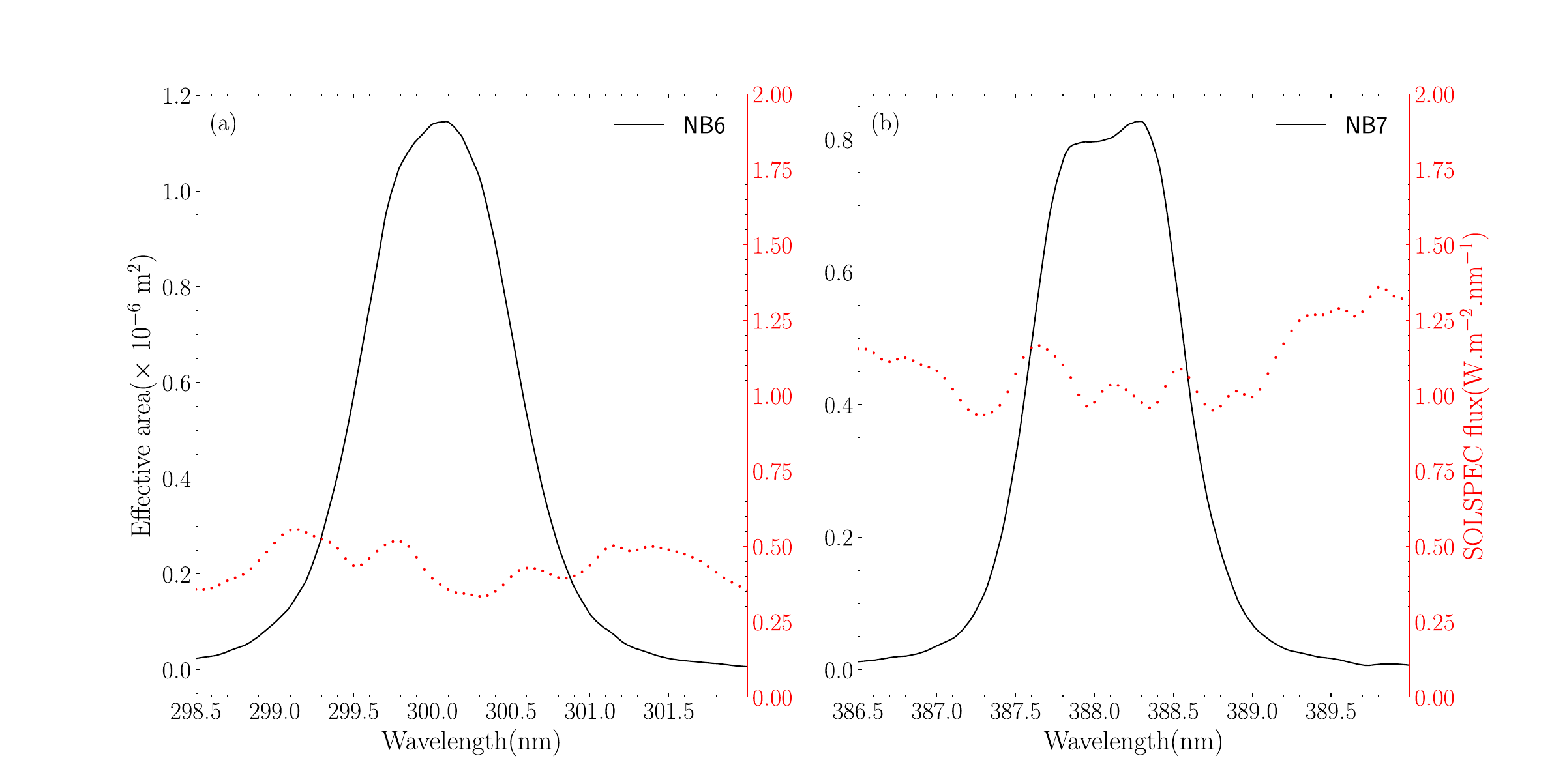}
    \caption{Spectral transmission profile of \suit in (a) NB06 and (b) NB07 bands with solar spectrum from SOLSPEC data.}
    \label{fig:nb7_solspec}
\end{figure}

To test this hypothesis, the lower-resolution SOLSPEC data is used to model the throughput for the NB06 filter. This is compared to that calculated for NB06 using higher-resolution SOLSTICE data. 
Fig. \ref{fig:nb7_solspec} shows the fluctuating nature of the sparsely populated SOLSPEC spectrum in continuum regions, along with the effective bandpasses of SUIT in NB06 and NB07 filters. The red dotted line represents the SOLSPEC solar spectrum, while the black solid lines represent the filter throughput. Due to sparse sampling of the solar spectrum, the throughput estimation is off from the measured values.
From Table \ref{tab:throughput}, we see that the ratio of measured vs. modeled throughput for NB06 using higher resolution SOLSTICE data is \js{1.07}. However, when the NB06 throughput is modeled using lower-resolution SOLSPEC data, the ratio obtained is $\mathrm{\left(\nicefrac{DN_{measured}}{DN_{throughput}}\right) \sim \js{1.96}}$. This result is similar to what we see for NB07 throughput modeled with SOLSPEC data. Therefore, the high ratio for NB07 in Table \ref{tab:throughput} can be attributed to the sparsely populated SOLSPEC data used to model the throughput.

\begin{table}[ht]
\caption{Comparison of the data numbers derived from throughput model (using SOLSTICE and SOLSPEC data) with the data numbers inferred from the measurements.} 
\label{tab:throughput}
\begin{center}
\begin{tabular}{|||c|c|c|c|||}
\hline
Science & Counts calculated & Counts calculated & Ratio of \\
filter & from measurement & from throughput model &  \\
 & $DN_{measured}~=~E_{\odot}\times~F$ & $DN_{throughput}~=~\int~P(\lambda)~R(\lambda)~t~d\lambda$ & $\frac{DN_{measured}}{DN_{throughput}}$\\
 & ($Data~number~s^{-1}$) & ($Data~number~s^{-1}$) & \\
\hline
NB02 & \js{57763} & 49099  & \js{1.18}\\
NB03 & \js{20142} & 15520  & \js{1.3}\\
NB04 & \js{38087} & 36717  & \js{1.04}\\
NB05 & \js{134909}& 131625 & \js{1.02}\\
NB06 & \js{114280}& 107064 & \js{1.07}\\
NB07 & \js{816026}& 391364 & \js{2.09}\\
NB08 & \js{28829} & 24096  & \js{1.2}\\
BB03 & \js{556220}& 434532 & \js{1.28}\\
\hline
\end{tabular}
\end{center}
\end{table}

\section{Full Payload Spectral Validation}\label{valid}
	
Spectral validation aims to verify the \suit payload's transmission profile with various science filters at different wavelengths within the respective filter bandpasses. Light from the monochromator is collimated and fed into the payload as before. The wavelength band of the source is chosen so that multiple measurements can be made at non-overlapping wavelength bins within the transmission band of a given filter. A holographic grating of 2400~lpmm blazed at 220~nm is used in the monochromator. The nominal dispersion of this grating is 0.74~nm/mm. The entrance slit width of the monochromator is set at 200~microns, providing a wavelength bin of \rv{0.148 nm}. This light is fed into the payload, and readings are taken across the transmission spectrum of the science filter, keeping adjacent readings independent of each other. The total energy is counted at each measured wavelength, and the spectral response of the entire telescope for the concerned filters is measured.
	
Following this, the entrance slit width of the monochromator is increased to 3~mm, and the light output is measured from the clear aperture of the collimator with the NIST traceable photodiode to verify the spectral flux being fed into the SUIT payload. This information is used to get a photometric calibration to the measured spectral flux. The spectral validation is performed with the setup illustrated in Fig.~\ref{fig:experimentalsetup}. \rv{Similar to the method described in \S\ref{valid}, we can estimate the spectral response of the concerned filters using the,

\begin{align}
    \mathrm{R(\lambda~=~\lambda_{1})~=~\frac{C(\lambda~=~\lambda_{1})}{E_{Photodiode}(\lambda~=~\lambda_{1})}~where,}
\end{align}

\begin{align*}
    \mathrm{R(\lambda~=~\lambda_{1})}~=&~\mathrm{The~spectral~response~of~a~given~filter~at}\\ 
    &~\mathrm{\lambda~=~\lambda_{1}(in~units~of~\nicefrac{DN.s^{-1}}{nW.m^{-2}.nm^{-1}})} \\
    \mathrm{C(\lambda~=~\lambda_{1})}~=&~\mathrm{Data~number~per~second~within~99\%~enclosed~counts}\\
    &~\mathrm{radius~of~the~fiber~image~at~\lambda~=~\lambda_{1}(in~units~of~DN.s^{-1})}\\
    \mathrm{E_{Photodiode}(\lambda~=~\lambda_{1})}~=&~\mathrm{Measured~input~spectral~irradiance~from~the~photodiode}\\
    &~\mathrm{at~\lambda~=~\lambda_{1}(in~units~of~nW.m^{-2}.nm^{-1})}
\end{align*}
}
 \begin{itemize}
 \rv{\item Similar to image acquisition during photometric calibration - the fiber face is opened and closed consecutively to record light from the data and the contribution from the background at various wavelengths within the concerned filter passband.
 
 \item The median value of the background is subtracted and the total counts within the 99\% enclosed energy radius ($C(\lambda_{1})$) is calculated in units of $\mathrm{DN.s^{-1}}$.
 
 \item The corresponding spectral irradiance is derived by dividing the photo-diode measurement ($E_{Photodiode}(\lambda_{1})$) ($\mathrm{nW.m^{-2}.nm^{-1}}$) with the spectral bin size.
 
 \item This gives us the response of the concerned filter at $\lambda~=~\lambda_{1}$ with, $\mathrm{R(\lambda_{1})~=~\frac{C(\lambda_{1})}{E_{Photodiode}(\lambda_{1})}}$\\$\mathrm{(in~units~of~\nicefrac{DN.s^{-1}}{nW.m^{-2}.nm^{-1}})}$ 
 }
 \end{itemize}

\begin{figure*}
\begin{center}
\begin{tabular}{c}
\includegraphics[trim={0cm 0cm 0cm 0cm},clip,width=0.9\linewidth]{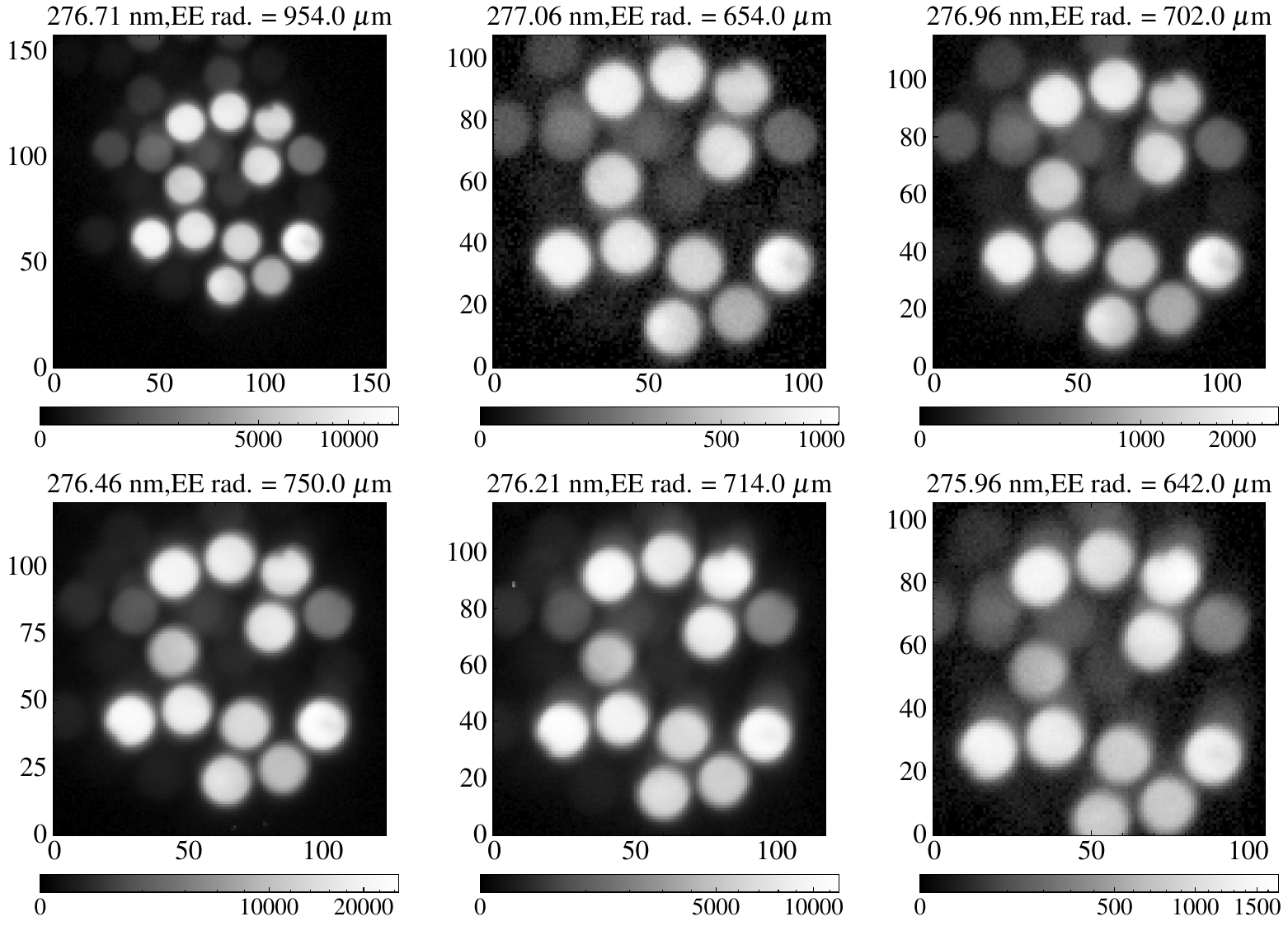}
\end{tabular}
\end{center}
\caption 
{\label{fig:nb2_images} Images captured at various wavelengths for the NB02 filter of \suit. The measurement wavelength and the 99\% \rv{encircled} energy radius are quoted at the top of each panel. \rv{The axes of the images are in pixels and the colorbar represents Data Numbers (DN).}} 
\end{figure*}

\begin{figure*}
\begin{center}
\begin{tabular}{c}
\includegraphics[trim={2.1cm 3.cm 1.2cm 5.2cm},clip,width=0.95\textwidth] {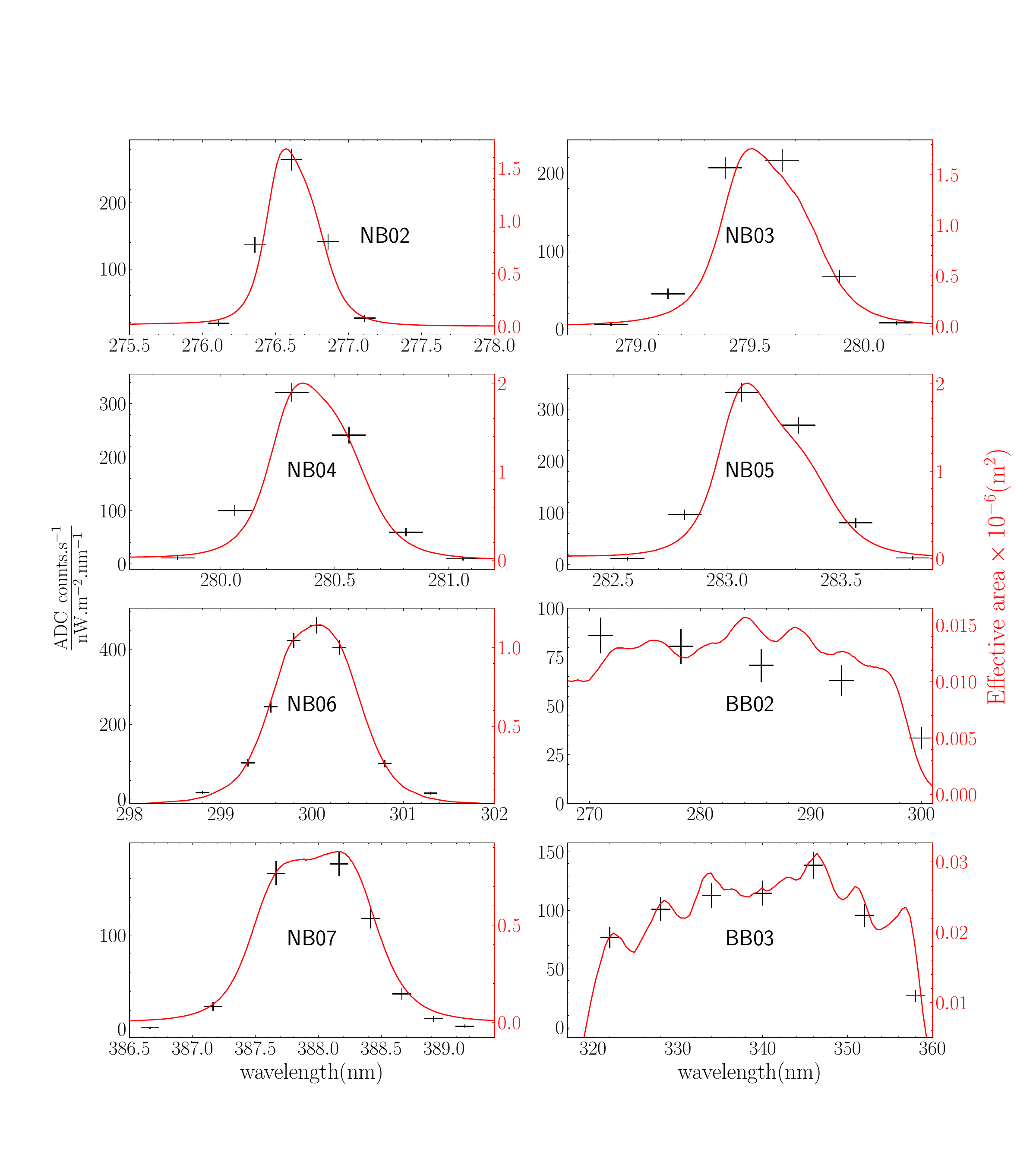}
\end{tabular}
\end{center}
\caption 
{\label{fig:sepc_calib} \js{Spectral validation for various filter combinations inferred from the imaging measurements (\rv{black} markers). The x-axis denotes the wavelength in nm and the y-axis denotes the ratio of ADC counts collected on the CCD with the spectral irradiance measured by the photodiode. The wavelength bins give the $x$ errors from the given input slit size. The Poisson uncertainty of the measured ADC counts gives the $y$ errors. The red solid curve shows the effective area calculated from measured transmission profiles of each component using Eqn \ref{eq:eff_area}.}}
\end{figure*}

In Fig.~\ref{fig:nb2_images}, we show the captured images for the NB02 filter combination at various wavelengths of incident light. The same was performed for all other filter combinations, but is not shown here for brevity. The measurement wavelength and the calculated 99\% \rv{encircled} energy radius are quoted at the top of each panel. We plot the spectral validation results of all filter combinations with transmission bandpasses above 250 nm in Fig.~\ref{fig:sepc_calib}. The \textit{x} and \textit{y} error bars in Fig.~\ref{fig:sepc_calib} represent wavelength bins for the given input slit size and the Poisson uncertainty of the measured ADC counts, respectively. \rv{The Poission uncertainty is given by the square root of the total detected photoelectrons in the image of an optical fiber. This is scaled to appropriate units while plotting.}

The plots demonstrate that the measured shape of the instrument spectral response agrees well with the modeled spectral response for all filter combinations with transmission bandpass above 250 nm. 

\begin{figure}
    \centering
    \includegraphics[width=0.8\textwidth]{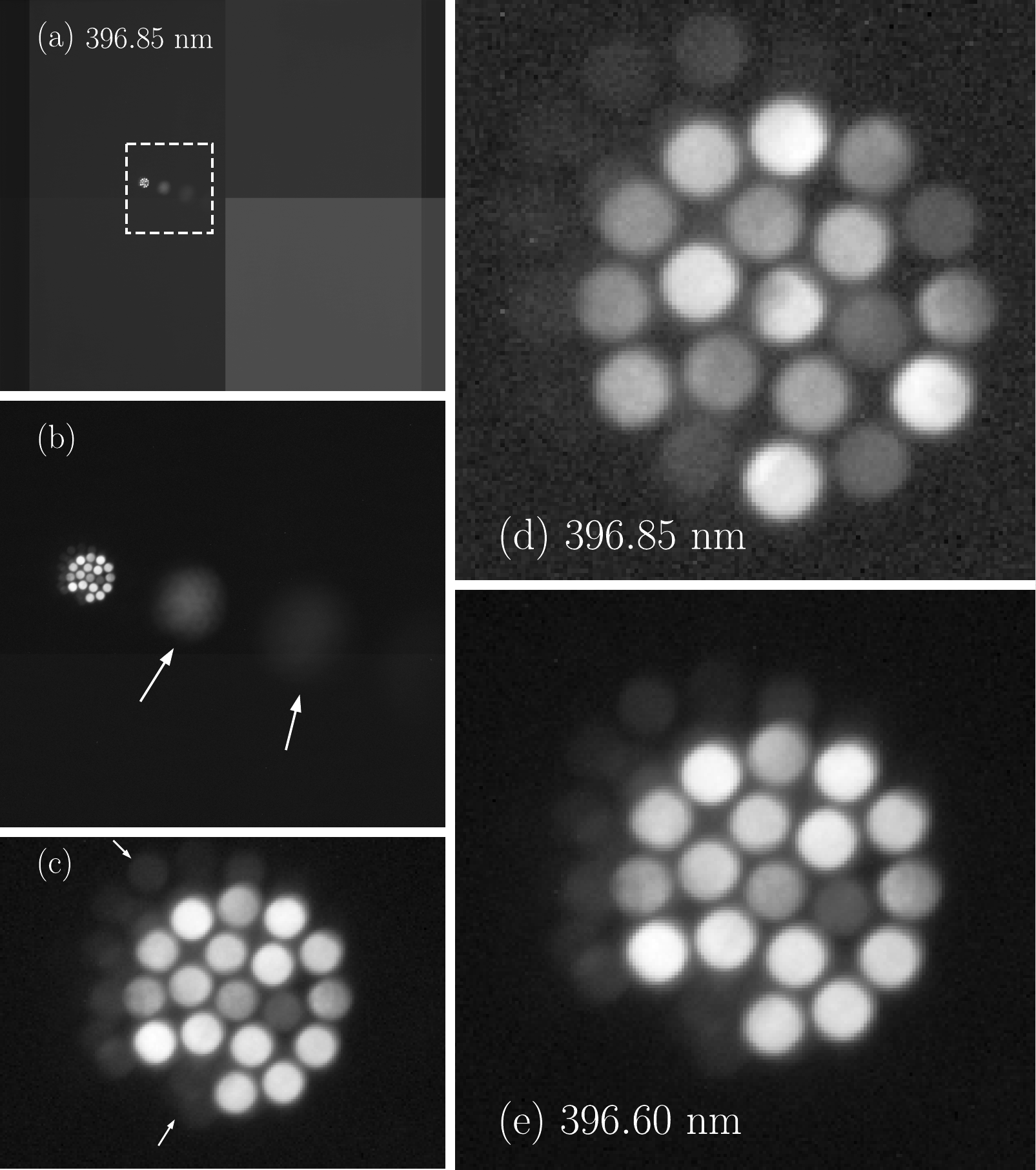}
    \caption{(a) The ghost reflections of the optical fiber as seen with the NB08 filter. The input beam is centered at 396.85 nm. The position of the ghost reflections is marked with a white dashed box. (b) The zoomed-in view of the white dashed box. The arrows indicate sparsely spaced ghost reflections. (c) The arrows indicate two closely spaced ghost reflections close to the center. (d) 99\% \rv{encircled} energy box around the fiber bundle at 396.85 nm. (e) 99\% \rv{encircled} energy box around the fiber bundle at 396.60 nm.}
    \label{fig:nb8_ghost}
\end{figure}

Each SUIT filter is mounted with a carefully chosen tilt, eliminating the possibility of inter-filter reflections as in Table \ref{table:tilt}. This is with the exception of NB08, where both filters are mounted without any relative angle between them, unlike other filter combinations (relative tilt of several degrees between filters). This filter has a 0.1 nm bandpass, and even slight tilts can cause a significant relative wavelength shift, potentially moving the Ca II h line out of the filter's in-band spectrum \cite{scfilt}.
Therefore, we could not carry out the spectral characterization for the NB08 filter due to the appearance of multiple strong ghost reflections close to the central patch.
In Fig. \ref{fig:nb8_ghost} (a), we plot the fiber image captured at 396.85~nm. The position of the central image of the fiber, along with the sparsely spaced ghost reflections, is marked with a white dashed box. Fig.~\ref{fig:nb8_ghost} (b) shows the zoomed-in view of the box. The sparsely shaped ghost reflections are marked with white arrows. Some ghost reflections are also seen very close to the central patch. Fig. \ref{fig:nb8_ghost} (c) shows a zoomed-in view around the central patch, with arrows indicating two closely spaced ghost reflections.
These ghost reflections appear very close to the central patch, and the relative brightness of the central patch to the ghost reflections changes as a function of wavelength, affecting the 99\% \rv{encircled} energy radius estimation as seen in Fig. \ref{fig:nb8_ghost} (d) and (e).

\begin{table}
    \centering
    \begin{tabular}{c c}
        \hline
        \textbf{Filter} & \textbf{Applied Tilt (deg)}\\ 
        \hline
        NB01 & 6 \\ 
        NB02 & 6 \\ 
        NB03 & 5 \\ 
        NB04 & 5 \\ 
        NB05 & 5 \\ 
        NB06 & 4 \\ 
        NB07 & 4 \\ 
        NB08\footnotemark[1] & 0 \\ 
        NB08\footnotemark[2] & 0 \\
        BB01 & 4 \\
        BB01 & 4 \\
        BB02 & 6 \\
        BB03 & 4 \\
        BP02 & 6 \\
        BP03 & 5 \\
        BP04 & 4 \\
        \hline
    \end{tabular}
    \footnotetext[1]{Mounted on filter wheel 1} 
    \footnotetext[2]{Mounted on filter wheel 2}
    \vspace{0.02\linewidth}
	\caption{\rv{Tilt angles of mounting science filters on \suit filter wheels. The columns indicate the filter name [col 1], and the applied angle of tilt [col 2].}}
	\label{table:tilt}
\end{table}
\section{Discussion \& Conclusion}\label{sec:conc}
In this document, we present the photometric calibration and full payload spectral validation for the \suit payload of Aditya-L1. The experimental setup comprises the payload maintained in an ultra-clean high vacuum environment and a collimator feeding light from a monochromator into the payload. The collimator must have high UV reflectance with comparable optical quality to the payload. Therefore, the flight spare model of \suit is used for this purpose.

The intensity of collimated light within a specific wavelength band is measured with a NIST-traceable photodiode and fed into the payload. These values are compared with forward-modeled throughput derived using sun-as-a-star spectrum from SOLSTICE and SOLSPEC. 
Measurements are made for filters with bandpasses above 250 nm.
The ratio of measured photoelectrons to that derived from the model is within \js {10\% for NB04, NB05 and NB06 filters, within 20\% for NB02 and NB08 filters, and within 30\% for NB03 and BB03 filters.}

\js {The discrepancy observed in NB07 at 388~nm (see Table~\ref{tab:throughput})} is due to the sparse resolution of SOLSPEC data, which is used to model throughputs above 310 nm. A comparative analysis of the improved modeling accuracy of SOLSTICE data as compared to SOLSPEC is also demonstrated.

The above calibration is not performed for NB01, BB01, and BB02 filters as the bandpasses are below 250 nm. These wavelengths get attenuated due to $\mathrm{O_3}$ formation in the atmosphere. 
Moreover, the Xenon light source transmission increases linearly by $\sim$ 60 times between 200 nm and 300 nm, beyond which the transmission is relatively flat till 400 nm. Thus, the data gathered for wavelengths below 250 nm has low SNR and is unreliable for photometry. 

\js{Each filter in the \suit science filter combinations studied here has transmission outside the bandpasses mentioned in Table \ref{tab:science_filters} in the 0.01\% to 0.001\% range. This prevents any out-of-band light originating anywhere outside the payload from reaching the CCD \cite{scfilt}}. Therefore, the disagreement between the experimental results and the model is not due to spectral leakage.

For the spectral validation, the wavelength bin of the input light is reduced compared to that during photometric characterization, and the response is measured at various wavelengths across the transmission bandpass of the filters. It is seen that the spectral response of the instrument matches the wavelength bands \suit is designed to observe (see Fig.\ref{fig:sepc_calib}). We could not perform the spectral validation for the NB08 filter due to multiple ghost reflections close to the central patch.

The payload performance measured in the lab agrees well with the modeled results, validating \suit's optical performance and presenting the reliability of the developed throughput model.

\subsection* {Code, Data, and Materials Availability} 
The data presented in this article are publicly available at \href{https://suit.iucaa.in/photometry_pub}{suit.iucaa.in/photometry\_pub}. Throughput modeling was performed using publicly available data from- SOLSPEC\cite{meftah_2018, meftah_2018_data} and SORCE SOLSTICE version 16 (\href{https://lasp.colorado.edu/sorce/data/}{https://lasp.colorado.edu/sorce/data/}).

\subsection* {Acknowledgments}
\rv{We thank the reviewers for their constructive comments and suggestions. \suit is built by a consortium led by the Inter-University Centre for Astronomy and Astrophysics (IUCAA), Pune, and supported by ISRO as part of the Aditya-L1 mission. The consortium consists of SAG/URSC, MAHE, CESSI-IISER Kolkata (MoE), IIA, MPS, USO/PRL, and Tezpur University. Aditya-L1 is an observatory class mission that is funded and operated by the Indian Space Research Organization. The mission was conceived and realised with the help from all ISRO Centres and payloads were realised by the payload PI Institutes in close collaboration with ISRO and many other national institutes - Indian Institute of Astrophysics (IIA); Inter-University Centre of Astronomy and Astrophysics (IUCAA); Laboratory for Electro-optics System (LEOS) of ISRO; Physical Research Laboratory (PRL); U R Rao Satellite Centre of ISRO; Vikram Sarabhai Space Centre (VSSC) of ISRO. This research has made use of the VizieR catalog access tool, CDS, Strasbourg, France (DOI: \href{10.26093/cds/vizier}{10.26093/cds/vizier}). The original description of the VizieR service was published in 2000, A\&AS 143, 23 \cite{vizier}. The results presented in this document also rely on data measured from the Solar Radiation Climate Experiment (SORCE) and are available at \noindent\href{https://lasp.colorado.edu/sorce/data/}{https://lasp.colorado.edu\\/sorce/data/}. These data were accessed via the LASP Interactive Solar Irradiance Datacenter (LISIRD) (\href{https://lasp.colorado.edu/lisird/}{https://lasp.colorado.edu/lisird/}). This research used PYTHON packages NumPy \cite{numpy}, Matplotlib \cite{matpltolib}, SciPy \cite{scipy}, and SciencePlots \cite{SciencePlots}. The 3D illustrations were made in Onshape \cite{onshape}.}

\subsection*{Disclosures}
The authors declare there are no financial interests, commercial affiliations, or other potential conflicts of interest that have influenced the objectivity of this research or the writing of this paper.

\bibliography{second_review_article}   
\bibliographystyle{spiejour}   

\vspace{1ex}
\noindent \textbf{Janmejoy Sarkar} is a research scholar at Tezpur University, India and a Senior Research Fellow at the Inter-University Centre for Astronomy and Astrophysics, India. He was instrumental in the mechanical integration, optical assembly and complete optical testing and calibration of the \suit payload on board Aditya-L1. He completed Masters in Physics with specialization in Astronomy and Astrophysics at Tezpur University, India. 

\noindent \textbf{Soumya Roy} is a research scholar at Inter-University Centre for Astronomy and Astrophysics, India. He was involved in the throughput modeling and characterizing the imaging performance and calibration of the \suit payload on board Aditya-L1. He completed his Masters in Physics from Presidency University, Kolkata, India.

\noindent Biographies and photographs of the other authors are not available.

\listoffigures
\listoftables

\end{document}